\documentclass[journal]{IEEEtran}

\usepackage{url}

\usepackage{breakurl}
\usepackage[breaklinks]{}

\usepackage{booktabs} % For formal tables
\usepackage{graphicx}
\usepackage{amsmath}
\usepackage{textcomp}
\usepackage{xcolor}
\usepackage{enumitem}
\usepackage{amssymb}% http://ctan.org/pkg/amssymb
\usepackage{pifont}% http://ctan.org/pkg/pifont
\usepackage{xspace} %MM
\usepackage{balance} %MM

\newcommand{\rc}{\textcolor{red}{\ding{51}}}%
\newcommand{\gc}{\textcolor{green}{\ding{51}}}%
\newcommand{\rx}{\textcolor{red}{\ding{55}}}%
\newcommand{\ra}{\textcolor{red}{\ding{217}}}%

\newcommand{\subhead}[1]{\vspace {1pt}\noindent{\textbf{#1.}}}

\newcommand{\red}[1]{\textcolor{red}{#1}}

\newcommand\blfootnote[1]{%
  \begingroup
  \renewcommand\thefootnote{}\footnote{#1}%
  \addtocounter{footnote}{-1}%
  \endgroup
}

\begin{document}
\title{The Sorry State of TLS Security in Enterprise Interception Appliances}

\author{Louis Waked, Mohammad Mannan, and Amr Youssef
\thanks{L. Waked, M. Mannan, and A. Youssef are with Concordia Institute for Information Systems Engineering, Concordia University, Montreal, Canada}
}

% \author{Michael~Shell,~\IEEEmembership{Member,~IEEE,}
%         John~Doe,~\IEEEmembership{Fellow,~OSA,}
%         and~Jane~Doe,~\IEEEmembership{Life~Fellow,~IEEE}% <-this % stops a space
% \thanks{M. Shell was with the Department
% of Electrical and Computer Engineering, Georgia Institute of Technology, Atlanta,
% GA, 30332 USA e-mail: (see http://www.michaelshell.org/contact.html).}% <-this % stops a space


\maketitle

\begin{abstract}

Network traffic inspection, including TLS traffic, in enterprise environments is widely practiced. Reasons for doing so are primarily related to improving enterprise security (e.g., phishing and malicious traffic detection) and meeting legal requirements (e.g., preventing unauthorized data leakage and copyright violations). To analyze TLS-encrypted data, network appliances implement a Man-in-the-Middle TLS proxy, by acting as the intended web server to a requesting client (e.g., a browser), and acting as the client to the actual/outside web server. As such, the TLS proxy must implement both a TLS client and a server, and handle a large amount of traffic, preferably, in real-time. However, as protocol and implementation layer vulnerabilities in TLS/HTTPS are quite frequent, these proxies must be, at least, as secure as a modern, up-to-date web browser, and a properly configured web server (e.g., an A+ rating in SSLlabs.com). As opposed to client-end TLS proxies (e.g., as in several anti-virus products), the proxies in network appliances may serve hundreds to thousands of clients, and \emph{any} vulnerability in their TLS implementations can significantly downgrade enterprise security.

To analyze TLS security of network appliances, we develop a comprehensive framework, by combining and extending tests from existing work on client-end and network-based interception studies. We analyze thirteen representative network appliances over a period of more than a year (including versions before and after notifying affected vendors, a total of 17 versions), and uncover several security issues. For instance, we found that four appliances perform no certificate validation at all, three use pre-generated certificates, and eleven accept certificates signed using MD5, exposing their clients to MITM attacks. Our goal is to highlight the risks introduced by widely-used TLS proxies in enterprise and government environments, potentially affecting many systems hosting security, privacy, and financially sensitive data.

\end{abstract}


\IEEEpeerreviewmaketitle

\section{Introduction}
\label{intro}
\blfootnote{This work is the extension of an ACM ASIACCS 2018 paper~\cite{waked}. The new contributions
are summarized under “Differences with the ACM ASIACCS version” in Section~\ref{intro}.}

\IEEEPARstart{M}{ost} network appliances currently include an SSL/TLS interception feature in their products. The interception process is performed by making use of a TLS web proxy server. Being either transparent or explicit to the end-user, the proxy intercepts the user's request to visit a TLS server, and creates two separate TLS connections. It acts as the HTTPS endpoint for the user's browser, and as the client for the actual HTTPS web server. Having the appropriate private key for the signing certificate (inserted to the client's root CA store), the proxy has access to the raw plaintext traffic, and can perform any desired action, such as restricting the access to the web page by parsing its content, or passing it to an anti-virus/malware analysis module or a customized traffic monitoring tool. Common reasons for adopting TLS interception include the protection of organization and individuals against malware and phishing attacks, law enforcement and surveillance, access control and web filtering, national security, hacking and spying, and privacy and identity theft~\cite{ruoti2016user}. 

While interception violates the implicit end-to-end guarantee of TLS, we focus on the potential vulnerabilities that such feature introduces to end-users located behind the network appliances, following several other existing studies on TLS interception, e.g.,~\cite{jarmoc2012ssl, WDormann, WDormann2, de2016killed, durumeric2017security}. In general, TLS interception, even if implemented correctly, still increases the attack surface on TLS due to the introduction of an additional TLS client and server at the proxy. However, the lack of consideration for following the current best practices on TLS security as implemented in modern browsers and TLS servers, may result in severe potential vulnerabilities, and overall, a significantly weak TLS connection.  For example, the proxy may not mirror the TLS version and certificate parameters or might accept outdated, insecure ones. Also, the proxy could allow TLS compression, enabling the CRIME attack~\cite{duong2012crime}, or insecure renegotiation~\cite{rescorla2010rfc}. The proxy may downgrade the Extended Validation (EV) domains to Domain Validated (DV) ones. It also may not mirror the cipher suites offered by the requesting client, and use a hard-coded list with weak and insecure ciphers, reviving old attacks such as FREAK~\cite{durumeric2015tracking}, Logjam~\cite{adrian2015imperfect}, and BEAST~\cite{duong2011here}. If the proxy does not implement a proper certificate validation mechanism, invalid and tampered certificates could be accepted by the proxy, and the clients (as they see only proxy-issued, valid certificates). Accepting its own root certificate as the signing authority of externally delivered content could allow  MITM attacks on the network appliance itself. The use of a pre-generated key pair by a proxy could enable a generic trivial MITM attack~\cite{de2016killed}. In addition, the proxy may rely on an outdated root CA store for certificate validation, containing certificates with insecure key length, expired certificates, or banned certificates that are no longer trusted by major browsers/OS vendors.

Concerns about security weaknesses introduced by TLS interception proxies are not new. In 2012, Jarmoc~\cite{jarmoc2012ssl} proposed a basic framework for testing network appliances consisting of seven certificate validation tests, and applied it on four network appliances. Dormann~\cite{WDormann, WDormann2} relied on badssl.com's tests to analyze the certificate validation process of two network appliances, revealing flaws in the appliances' certificate validation mechanisms.
Carnavalet and Mannan~\cite{de2016killed} designed a framework for analyzing client-based TLS proxies (as included in several leading anti-virus and parental control applications), and revealed several flaws in the TLS version and certificate mapping, certificate validation, private key generation and protection, CA trusted store content, in addition to vulnerabilities to known TLS attacks. In 2017, Durumeric et al.~\cite{durumeric2017security} applied tests from earlier frameworks on twelve network appliances and thirteen client-side TLS proxies, uncovering several flaws in certificate validation, cipher suites, TLS versions and known TLS attacks. 

We argue that most past studies on network appliances analyzed only preliminary aspects of TLS interception, while the extensive work of Carnavalet and Mannan~\cite{de2016killed} targeted only client-end TLS proxies. However, TLS vulnerabilities in network appliances could result in more serious security issues, as arguably, enterprise computers handle more important business/government data in bulk, compared to personal information on a home user machine. Also, a single, flawed enterprise TLS proxy can affect hundreds of business users, as opposed to one or few users using a home computer with a client-side TLS proxy.

We present an extensive framework dedicated for analyzing TLS intercepting appliances, borrowing/adapting several aspects of existing work on network appliances and client-end proxies, in addition to applying a set of comprehensive certificate validation tests. We analyze the TLS-related behaviors of appliance-based proxies, and their potential vulnerabilities from several perspectives: TLS version and certificate parameter mapping, cipher suites, private key generation/protection, content of root CA store, known TLS attacks, and 32 certificate validation tests. We use this framework to evaluate 13 representative TLS network appliances, a total of 17 product versions, between July 2017 and March 2018 (see Table~\ref{table6}), including open source, free, low-end, and high-end network appliances, and present the vulnerabilities and flaws found. All our findings have been disclosed to the respective companies. 

A summary of our findings include the following. Four appliances do not perform \emph{any} certificate validation by default, allowing simple MITM attacks against their clients; one does not perform certificate validation even after explicitly enabling this feature in its configuration. Another appliance accepts self-signed certificates, and three appliances use pre-generated key pairs, enabling similar MITM attacks. One appliance states in its documentation that it generates the key pair during installation, but we found it to use a pre-generated key pair. Four appliances accept their own certificates for externally delivered content. Eleven appliances accept certificates signed using the MD5 algorithm, and four appliances accept certificates signed using the MD4 algorithm. Eight appliances offer weak and insecure ciphers (generally not offered by any modern browser). Four appliances support SSL 3.0, of which one only accepts TLS 1.0 and SSL 3.0 (i.e., rejects connections with TLS 1.1 and 1.2). We also found that the root CA stores of all appliances include at least one or more certificates deemed untrusted by major browser/OS vendors, and one appliance includes an RSA-512 certificate, which can be trivially compromised. Nine appliances also do not encrypt their private keys; seven such keys are accessible to unprivileged processes running on the same appliance.

Analyzing network appliances raises several new challenges compared to testing browsers and client-end TLS proxies. Several network appliances do not include an interface for importing custom certificates (essential for testing), and many appliances do not provide access to the file system or a terminal, overburdening the tasks of injecting custom certificates and locating the private keys (for details, see~\cite[Appendix B]{waked}). Many appliances do not support more than one or two network interfaces, and thus, require the use of a router to connect to multiple interfaces. In addition, appliances that perform SSL certificate caching require the generation of a new root key pair for their TLS proxies for each test.

Our contributions can be summarized as follows: (1) We develop a comprehensive framework to analyze TLS interception in enterprise-grade network appliances, combining our own certificate validation tests with existing tests for TLS proxies (both client-end services and network appliances), which we reuse or adapt as necessary for our purpose. Our certificate validation tests can be found at: \url{https://madiba.encs.concordia.ca/software/tls-netapp/}.
(2) We use this framework to evaluate thirteen well-known appliances from all tiers: open source, free, low-end, and high-end products, indicating that the proposed framework can be applied to different types of network appliances. (3) We uncover several vulnerabilities and bad practices in the analyzed appliances, including: either an incomplete or completely absent certificate validation process (resulting trivial MITM attacks), improper use of TLS parameters that mislead clients, inadequate private key protection, and the use of weak/insecure cipher suites. 

\subhead{Differences with the ACM ASIACCS version~\cite{waked}}
We make the following significant changes to the current article. We analyze seven new appliances using our framework (in addition to the six appliances in~\cite{waked}); the updated results are discussed in Section~\ref{results1}. In total, we report the results of seventeen different versions of thirteen products we analyzed. We uncover a new dangerous vulnerability that was not seen with the previously tested appliances; three of the newly tested appliances use pre-generated root key pairs. This can lead to trivial full server impersonation attacks; we discuss the implications of this vulnerability in Section~\ref{practicalattacks}. We also reanalyze the latest up-to-date releases of the six previously analyzed appliances, present the new results in Section~\ref{results1}, and discuss the differences between the findings of the previously tested releases and the new ones in Section~\ref{differences}. Note that we shared our findings with the affected product vendors before our initial publication, and tested the versions that some vendors claimed to have fixed the reported vulnerabilities.
We also make a few modifications to our original test framework (Section~\ref{framework1}): we improve the key/certificate extraction from the appliances (Section~\ref{catrusted}); we introduce a new method for locating private keys (Section~\ref{ownkey}) in some appliances using the `squid.conf' file, and brute-forcing the passphrase of the encrypted private key in one appliance using a password cracker tool; and we enumerate the list of TCP ports that need to be intercepted to successfully utilize the Qualys client test and badssl.com (Section~\ref{architecture}).

\begin{table}[tbp]
\centering
\caption{List of the tested appliances; we use the text in bold to refer to the appliances throughout this paper. We only discuss the results of the latest releases (``Versions'' in bold).}
\label{table6}
\scalebox{0.9}{
\begin{tabular}{l|l|l}
\hline
\textbf{Appliance} & \textbf{Company} & \textbf{Version} \\ \hline
\textbf{Untangle} NG Firewall &  Untangle & 13.0 \\
 &   & \textbf{13.2} \\
\textbf{pfSense} & NetGate & 2.3.4 \\
 &  & \textbf{2.4.2-P1} \\
\textbf{WebTitan} Gateway & TitanHQ & 5.15 build 794 \\
 &  & \textbf{5.16 build 1602} \\
\textbf{Microsoft} TMG & Microsoft & \textbf{2010 SP2 rollup update 5} \\
\textbf{UserGate} Web Filter & Entensys & \textbf{4.4.3320601} \\
\textbf{Cisco} Ironport WSA & Cisco & AsyncOS 10.5.1 build 270 \\
 &  & \textbf{AsyncOS 10.5.1 build 296} \\
\textbf{Sophos} UTM & Sophos & \textbf{9.506-2} \\
\textbf{TrendMicro} Interscan WSVA & TrendMicro & \textbf{6.5 SP2 build 1765} \\
\textbf{McAfee} Web Gateway & McAfee & \textbf{7.7.2.8.0 build 25114} \\
\textbf{Cacheguard} Web Gateway v1.3.5 & Cacheguard & \textbf{1.3.5} \\
\textbf{OpenSense} & Deciso B.V. & \textbf{18.1.2\_2} \\
\textbf{Comodo} Dome Firewall & Comodo & \textbf{2.3.0} \\
\textbf{Endian} Firewall Community & Endian & \textbf{3.2.4} \\
\hline
\end{tabular}}
%\vspace{-15pt}
\end{table}

\section{Background}\label{background1}
In this section, we describe the TLS interception process, list the tested products, state the expected behavior of a TLS proxy, and explain the threat model.

\subhead{Terminology}
Throughout the paper, we refer to the TLS intercepting network appliances as proxies, HTTPS proxies, TLS proxies, middleboxes, or simply appliances. For the TLS requesting client, we use: browser, end-user, user, or client. The term \emph{mirroring} is used to describe a situation where the proxy sends the same TLS parameters received from the web server to the client side, and vice versa; otherwise, \emph{mapping} is used to indicate that the proxy has modified some parameters (for better or worse). We refer to the trusted root CA stores as stores, trusted stores or trusted CA stores. We refer to virtual machines as virtual appliances, VMs, or simply machines.

%\vspace{-10pt}
\subsection{Proxies and TLS Interception}
For TLS interception, network appliances make use of TLS proxies, deployed as either transparent proxies or explicit proxies. The explicit proxy requires the client machine or browser to have the proxy's IP address and listening port specifically configured to operate. Thus, the client is aware of the interception process, as the requests are sent to the proxy's socket. On the other hand, transparent proxies may operate without the explicit awareness of the clients, as they intercept outgoing requests that are meant for the web servers, without the use of an explicit proxy configuration on the client side; however, for TLS interception, a proxy's certificate must be added to the client's trusted root CA store (explicitly by the end-user, or pre-configured by an administrator). Such proxies could filter all ports, or a specific set of ports, typically including HTTP port 80 and HTTPS port 443. Secure email protocols could also be intercepted, by filtering port 465 for secure SMTP, port 993 for secure IMAP, and port 995 for secure POP3. The proxy handles the client's outgoing request by acting as the TLS connection's endpoint, and simultaneously initiates a new TLS connection to the actual web server by acting as the client, while relaying the two connections' requests and responses.

By design, the TLS protocol should prevent any MITM interception attempt, by enforcing a certificate validation process, which mandates that the incoming server certificate must be signed by a trusted issuer. Certificate authorities only provide server certificates to validated domains, and not to forwarding proxies, precluding the proxy from becoming a trusted issuer (i.e., a \emph{valid} local CA). To bypass this restriction, the proxy can use a self-signed certificate that is added to the trusted root CA store of the TLS client, and thereby allowing the proxy to sign certificates for \emph{any} domain on-the-fly, and avoid triggering browser warnings that may expose the untrusted status of the proxy's certificate. 
Thereafter, all HTTPS pages at the client will be protected by the proxy's certificate, instead of the intended external web server's certificate. Users are not usually aware of the interception process, unless they manually check the server certificate's issuer chain, and notice that the issuer is a local CA~\cite{o2016tls}.

%\vspace{-8pt}
\subsection{Tested Appliances}
Most current network appliance vendors offer products for TLS interception. We select thirteen products, including: free appliances, appliances typically deployed by small companies, appliances with affordable licensing for small to medium sized businesses, and high-end products for large enterprises; see Table~\ref{table6}. On a side note, we performed several rounds of updates and patches for Microsoft, on a Windows Server 2008 R2 operating system, as recommended by Microsoft's documentation~\cite{TMGWind2008}. These include the service pack 1 (SP1), the service pack 1 update, the service pack 2, and five rollup updates (1 to 5)~\cite{TMG}.

For all the analyzed appliances, we keep the default configuration for their respective TLS proxies. An administrator could of course manually modify this default configuration, which may improve or degrade the proxy's TLS security. We thus choose to apply our test framework on the unmodified configuration (assuming the vendors will use \emph{secure-defaults}).

%\vspace{-8pt}
\subsection{Expected Behavior of a TLS Proxy}

We summarize expected behaviors from a prudent interception proxy (following~\cite{de2016killed}). Deviations from these behaviors help design and refine our framework and validation tests.

The TLS version, key length, and signature algorithms should be mirrored (between client-proxy and proxy-web) to avoid misleading clients regarding the TLS security parameters used in the proxy-web connection. The client's cipher suites also should be mirrored, or at least the proxy must not offer any weak ciphers. The proxy must be patched against known TLS attacks, e.g., BEAST~\cite{duong2011here}, CRIME~\cite{duong2012crime}, FREAK~\cite{durumeric2015tracking}, Logjam~\cite{adrian2015imperfect}, and TLS insecure renegotiation~\cite{rescorla2010rfc}.  

TLS proxies must properly validate external certificates, as the client software (e.g., browsers) will be only exposed to the proxy-issued certificates.  
Thus, all aspects of TLS chain of trust should be properly validated, and common flaws must be avoided, e.g., untrusted issuers, mismatched signatures, wrong common-names, constrained issuers, revoked/expired certificates, and deprecated signature algorithms. The proxy's trusted CA store must avoid short key, expired or untrusted issuer certificates. TLS proxies must also disallow any external certificates signed by their own root keys. 
Proxies' private keys must be adequately protected (e.g., limiting access to the root account), and the keys must not be pre-generated (cf.~\cite{rosenblatt2015lenovo}).

%\vspace{-8pt}
\subsection{Threat Model}
We mainly consider three types of attackers.

An \emph{external attacker} can impersonate any web server by performing a MITM attack on a network appliance that does not perform a proper certificate validation. The attacker could be anywhere on the network between the appliance and the target website. Even if the validation process is perfect, the attacker could still impersonate any web server, if the appliance uses a pre-generated root certificate or accepts external site-certificates signed by its own root key. The attacker could also take advantage of known TLS attacks/vulnerabilities to potentially acquire authentication cookies (BEAST, CRIME), or impersonate web servers (FREAK, Logjam).

A \emph{local attacker} (e.g., a malicious insider, external attacker with access to a compromised machine in the local network) with a network sniffer in promiscuous mode can get access to the raw traffic from the connections between the network appliance and clients. If the appliance uses a pre-generated certificate, the malicious user can install her own instance of the appliance, acquire its private key, and use it to decrypt the sniffed local traffic when the TLS connections are not protected by forward-secure ciphers. Such an adversary can also impersonate the proxy itself to other client machines, although this may be easily discovered by network administrators.

An \emph{attacker who compromises the network appliance} itself with non-root privileges can acquire the private key if the key is not properly protected (e.g., read access to `other' users and no passphrase encryption). With elevated privileges, more powerful attacks can be performed (e.g., beyond accessing/modifying TLS traffic). We do not consider such privileged attackers, assuming having root access on the appliance would be much more difficult than compromising other low-privileged accounts. Note that, in most cases, the \emph{appliance} is simply an ordinary Linux/Windows box with specialized software/kernel, resulting in a large trusted code base (TCB).

\section{Related Work}\label{related1}
Several studies have been recently conducted on TLS interception, TLS certificate validation, and forged TLS certificates. 

%
% \subhead{Comparison}
The most closely related work is by Durumeric et al.~\cite{durumeric2017security} (other studies mostly involved analyzing TLS libraries and client-end proxies; see~\cite{waked}). While their work focuses primarily on fingerprinting TLS interception, in addition to a brief security measurement for several HTTPS proxies, we develop an extensive framework dedicated for analyzing the TLS interception on network appliances. They checked/rated the highest TLS version  supported by a target proxy, while we examine all the supported versions by the proxy, in addition to their respective mapping/mirroring to the client side. Durumeric et al.'s certificate validation tests include: expired, self-signed, invalidly signed certificates, and certificates signed by CAs with known private keys; we include more tests for this important aspect (a total of 32 distinct tests). We also include several new tests such as: checking the content of the CA trusted store and the certificate parameter mapping, locating the private keys of the proxies and examining their security (including checking pre-generated root certificates); these tests are mostly added/extended from~\cite{de2016killed, chausymcerts}.
In terms of results, for the five products overlapping with Durumeric et al.~\cite{durumeric2017security}, we observed a few differences; see Section~\ref{differences}.  

\section{Proposed Framework}\label{framework1}

In this section, we present the setup/architecture of the proposed framework, and its major components and tests.

\begin{figure}
\centering 
\includegraphics[scale=1.2]{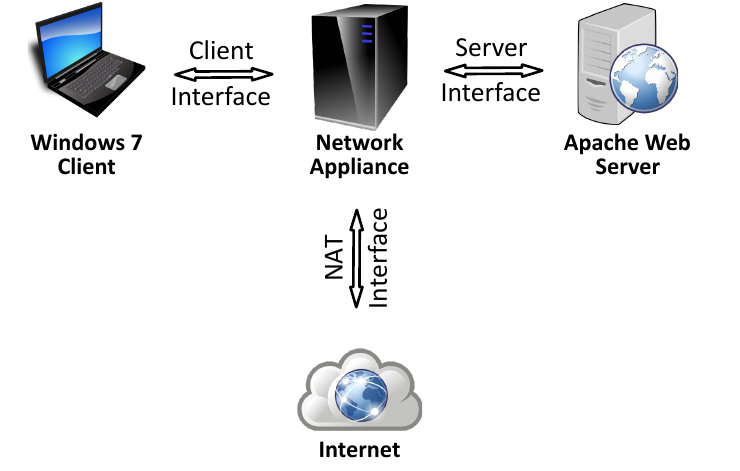}
\setlength{\belowcaptionskip}{-10pt}
\setlength{\abovecaptionskip}{-10pt}
%\vspace{-5pt}
\caption{Framework components and the overall test architecture}
\label{figure1}
%\vspace{-18pt}
\end{figure}

%\textbf{\large Framework Architecture}
%\vspace{-8pt}
\subsection{Test Setup/Architecture}
\label{architecture}
Our framework consists of three virtual machines: a client, a web server, and the TLS intercepting network appliance; see Figure~\ref{figure1}.
The client machine (Windows 7 SP1) is located behind the appliance; we update the client with all available Windows updates, and install up-to-date Mozilla Firefox, Google Chrome, and Internet Explorer 11 on it. We insert the TLS proxy's root certificate into the client's trusted CA stores (both Windows and Mozilla stores). We use a browser to initiate HTTPS requests to our local Apache web server, and the online TLS security testing suites (for certain tests). These requests are intercepted by the tested TLS proxy.

The second machine hosts a web server (Apache/Ubuntu 16.04), and accepts HTTP and HTTPS requests (on ports 80 and 443, respectively); all port 80 requests are redirected to port 443. It is initially configured to accept all TLS/SSL protocol versions, and all available cipher suites. The server name is configured to be \emph{apache.host}, as the crafted certificates must hold a domain name (not an IP address). We generate the faulty certificates using OpenSSL, which are served from the web server. It also hosts the patched \url{howsmyssl.com} code~\cite{jmhodges_2013}.

The pre-installed OpenSSL version on the Ubuntu 16.04 distribution is not compiled with SSLv3 support. Thus, in order to test the acceptance and mapping of SSLv3 only, we rely on an identically configured older version of Ubuntu (14.04), with an older OpenSSL version that supports SSLv3.

The third machine hosts the appliance that we want to test. The appliances are typically available as a trial version on a vendor's website, with a pre-configured OS, either as an ISO image or an Open Virtualization Format file. The appliances are configured to intercept TLS traffic either as a transparent or explicit proxy, depending on the available modules. If both are available, transparent proxies are prioritized, as they do not require any client-side network configuration. We disable services such as firewall and URL filtering, if bundled in the appliances, to avoid any potential interferences in our TLS analysis. The root CA certificates corresponding to our faulty test certificates are injected into the trusted stores of the appliances. We include the following TCP ports to the list of intercepted ports, as they are used by the Qualys client test and badssl.com: 1010, 1011, 10200, 10300, 10301, 10302, 10303, 10444, and 10445 (determined by analyzing traffic captures on Wireshark of TLS connections to Qualys/badssl websites). Some appliances offer an interface to add custom ports to be intercepted by the TLS proxy, while others require manual configuration in their configuration files.

\begin{figure} [t] % dropped "hb" => could be danger
\centering 
\includegraphics[scale=1.07]{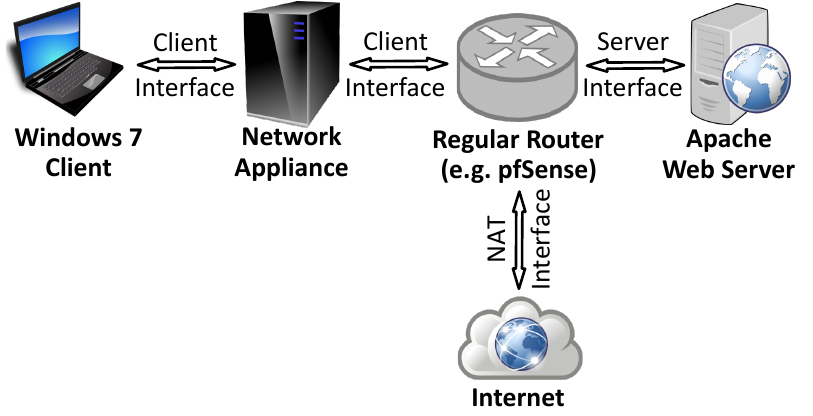}
\setlength{\belowcaptionskip}{-5pt}
\setlength{\abovecaptionskip}{-10pt}
\vspace{-5pt}
\caption{Framework components and test architecture with a router}
\label{figure2}
%\vspace{-15pt}
\end{figure}

We set up a local DNS entry for \emph{apache.host} on the client, web servers and network appliances machines. 
Operating systems match local DNS entries, found typically in the \emph{hosts} file, before remote DNS entries, resulting in the correct mapping of our test server's domain name to its corresponding IP address.

We require three different interfaces on each virtual network appliance. The \emph{Client Interface} is used to connect to the Windows 7 client. The traffic incoming from this interface is intercepted by the TLS proxy. Transparent proxies only require the appliance to be the default gateway for the client, while explicit proxies require the client to configure the proxy settings with the appliance's socket details. 
The \emph{Server Interface} is used to connect to the Apache web server. The \emph{WAN Interface} connects to the Internet, through Network Address Translation (NAT). However, some appliances support one or two interfaces. In such cases, we add a fourth virtual machine, that acts solely as a router with multiple interfaces. We use pfSense as the router (without TLS interception), relying on it for NATting and routing traffic of the three interfaces as required; see Figure~\ref{figure2}. The client and the network appliance are connected to the \emph{Client Interface}, the web server is connected to the \emph{Server Interface}, and the Internet connectivity is provided through NAT on a third interface via pfSense. A local DNS entry for \emph{apache.host} is also added to this router. 

%\vspace{-8pt}
\subsection{Trusted CA Store}\label{catrusted}

We first locate the trusted store of the TLS proxy, to inject our root certificates in it, required for most of our tests.

Injecting custom certificates into a trusted store could be trivial, if the appliance directly allows adding custom root CAs (e.g., via its user interface). If no such interface is offered, we attempt to get a command line (\emph{shell}) access through a terminal, or, the SSH service if available, by enabling the SSH server first through the settings panels (we transfer files using the SCP/SFTP protocols). 
If SSH is unavailable, we mount the virtual disk image of the appliance on a separate Linux machine. When mounting, we perform several attempts to find the correct filesystem type and subtype used by the appliance (undocumented). After a successful mount, we search the entire filesystem for digital certificates in known formats, such as ``.crt'', ``.pem'', ``.cer'', ``.der'', and ``.key''. We thus locate several directories with candidate certificates, and subsequently delete the content of each file, while trying to access regular websites from the client. When an ``untrusted issuer'' warning appears at the client, we then learn the exact location/directory of the trusted store (and can eliminate duplicate/unnecessary certificates found in multiple directories). 

We then inject our custom crafted root certificates into the trusted CA stores. We also parse the certificates available in the trusted stores to identify any expired certificates, or certificates with short key lengths (e.g., RSA-512 and RSA-1024). We also check for the presence of root CA certificates from issuers that are no longer trusted by major browser/OS vendors. Our list of misbehaving CAs includes:  China Internet Network Information Center (CNNIC~\cite{CNNIC}), T\"URKTRUST~\cite{ducklin2013turktrust}, ANSSI~\cite{ANSSI}, WoSign~\cite{wosign_smartcom}, Smartcom~\cite{wosign_smartcom}, and Diginotar~\cite{prins2011diginotar}.

%\vspace{-8pt}
\subsection{TLS Version Mapping}

To test the SSL/TLS version acceptance and TLS parameter mapping/mirroring, we alter the Apache web server's configuration. We use a valid certificate whose root CA certificate is imported into the trusted stores of the client (to avoid warnings and errors). We then subsequently force one TLS version after another at the web server, and visit the web server from the client, while documenting the versions observed in the browser's HTTPS connection information. Using this methodology, we are able to analyze the behavior of a proxy regarding each SSL/TLS version: if a given version is blocked, allowed, or altered in the client-to-proxy HTTPS connection.

%\vspace{-8pt}
\subsection{Certificate Parameters Mapping}
We check if the proxy-to-server certificate parameters are mapped or mirrored to the client-to-proxy certificate parameters. The parameters studied are signature hashing algorithms, certificate key lengths, and the EV/DV status.

For testing signature hashing algorithms, we craft multiple valid certificates with different \emph{secure} hash algorithms, such as SHA-256, SHA-384 and SHA-512. We import their root CA certificates into the trusted stores of the client to avoid warnings and errors. We subsequently load each certificate and its private key into the web server, and visit the web page from the browser. We track the signature  algorithms used in the certificates generated by the TLS proxy for each connection, and learn if the proxy mirrors the signature hashing algorithms, or use a single hard-coded one.

For testing certificate key lengths, we craft multiple certificates with multiple acceptable key sizes: RSA-2048, RSA-3072 and RSA-4096. We import their correspondent root CA certificates into the trusted stores of the client. We subsequently load each certificate and its private key into the web server, and visit the web page from the browser. We check the key length used for the client-to-proxy server certificate generated by the proxy for each connection, and learn if the proxy mirrors the  key-length, or uses a hard-coded length.

We rely on Twitter's website to study the network appliance's behavior regarding EV certificates. We visit \url{twitter.com} on the client machine, and check the client-to-proxy certificate displayed by the browser. TLS proxies can identify the presence of EV certificates (e.g., to avoid downgrading them to DV), by parsing the content and locating the CA/browser forum's EV OID: 2.23.140.1.1~\cite{wilson}.

%\vspace{-8pt}
\subsection{Cipher Suites}
Cipher suites offered by the TLS proxy in the proxy-to-server TLS connection can be examined in multiple ways. We initially rely on publicly hosted TLS testing suites, \url{howsmyssl.com} and the Qualys client test~\cite{qualys2014labs}. Since the connection is proxied, the displayed results found on the client's browser are the results of the proxy-to-server connection, and not the client-to-proxy connection. If the mentioned web pages are not filtered, for reasons such as the use unfiltered or non-standard ports, we use Wireshark to capture the TLS packets and inspect the Client Hello message initiated by the proxy to locate the list of ciphers offered.

We then compare the list of ciphers offered by the proxy to that list of our browsers, deduce if the TLS proxy performs a cipher suite mirroring or uses a hard-coded list. We also parse the proxy's cipher-suite for weak and insecure ciphers that could lead to insecure and vulnerable TLS connections.

%\vspace{-8pt}
\subsection{Known TLS Attacks}
We test TLS proxies for vulnerabilities against well-known TLS attacks, including: BEAST, CRIME, FREAK, Logjam, and Insecure Renegotiation.
We rely on the Qualys SSL Client Test~\cite{qualys2014labs} to confirm if the TLS proxy is patched against FREAK, Logjam, and Insecure Renegotiation, and check if TLS compression is enabled for possible CRIME attacks. We visit the web page from the client browser, which displays the results for the proxy-to-server TLS connection.
For the BEAST attack, we rely on \url{howsmyssl.com}~\cite{jmhodges_2013} (with the modifications from~\cite{de2016killed}) to test the proxies that support TLS 1.2 and 1.1. For a system to be vulnerable to BEAST~\cite{duong2011here}, it must support TLS 1.0, and use the CBC mode. However, after the BEAST attack was uncovered, a patch was released for CBC (implementing the $1/(n-1)$ split patch~\cite{su_2011}), but was identically named as CBC, making the distinction  between the patched/unpatched CBC difficult.

%\vspace{-8pt}
\subsection{Crafting Faulty Certificates}
We use OpenSSL to craft our invalid test certificates, specifying \emph{apache.host} as the Common Name (CN), except for the wrong CN test. We then deploy each certificate on our Apache web server, and request the HTTPS web page from the proxied client, and thus learn how the TLS proxy behaves when exposed to faulty certificates; if a connection is allowed, we consider the proxy is at fault. If the proxy replaces the faulty certificate with a valid one (generated by itself), leaving no way even for a prudent client (e.g., an up-to-date browser) to detect the faulty remote certificate, we consider this as a serious vulnerability. If the proxy passes the unmodified certificate and relies on client applications to react appropriately (e.g., showing warning/error messages, or terminating the connection), we still consider the proxy to be at fault for two reasons: (a) we do not see any justification for allowing plain, invalid certificates by any TLS agent, and (b) not all TLS client applications are as up-to-date as modern browsers, and thus may fail to detect the faulty certificates.

When the certificate's chain of trust contain intermediate certificate(s), we place the leaf certificate and intermediate certificate(s) at the web server, by appending the intermediate certificate(s) public keys after the server leaf certificate, in \emph{SSLCertificateFile}. Note that we inject the issuing CA certificates of the crafted certificates into the TLS proxy's trusted store for all tests, except for the unknown issuer test and the fake GeoTrust test.  

We compile the list of faulty certificates using several sources (including~\cite{de2016killed,chausymcerts,housley2008rfc}; see~\cite{waked}) for details).
Before using the faulty certificates, we assess them against Firefox v53.0 (latest at the time of testing), and confirm that Firefox terminates all connections with these certificates.

As part of the analysis of the certificate validation mechanisms, we ensure that the TLS proxies do not cache TLS certificates, by checking the `Organization Name' field of the subject parameter in the server certificates. Each leaf certificate of the crafted chains contains a unique `Organization Name' value, allowing us to identify exactly which TLS certificate is being proxied. We additionally check if the TLS inspection feature is enabled by default after the activation of the appliances, or if it requires a manual activation.

%Table 1
%\hspace{0.05cm}
\begin{table*}[htb]
\centering
\caption{Results for TLS parameter mapping/mirroring and vulnerabilities to known attacks. For TLS version mapping, we display the TLS versions seen by the client when the web server uses TLS 1.2, 1.1, 1.0 and SSL 3.0 (`--' means unsupported. `\textdagger' means supported but terminate with a handshake failure; see Section~\ref{tlspararesults}). Under ``Key Length Mapping'': `*' means the appliance mirrors RSA-512 and RSA-1024 key sizes, but use a static key size RSA-2048 for any higher key sizes (see Section~\ref{tlsversionmapping}). Under ``Problematic Ciphers'': Weak means deprecated; Insecure means broken; blank means good ciphers. Under ``BEAST'': \rx \textcolor{black}{} means vulnerable; \rx\red{*} \color{black}{}means potentially vulnerable (unknown if CBC is patched with $1/(n-1)$ split); blank means patched. All the appliances are patched against FREAK, Logjam, CRIME, and Insecure Renegotiation.}
\label{table1}
\resizebox{\textwidth}{!}{%
\begin{tabular}{c|c|c|c|c|c|c|c|c|c|c|c|c|c|c|}
\cline{2-11}
 & \multicolumn{4}{c|}{TLS Version Mapping} & \multicolumn{2}{c|}{Cipher Suites} & \multicolumn{3}{c|}{Certificate Parameter Mapping} &  \\ \cline{2-10} 
 & \begin{tabular}[c]{@{}l@{}}TLS\\ 1.2\end{tabular} & \begin{tabular}[c]{@{}l@{}}TLS\\ 1.1\end{tabular} & \begin{tabular}[c]{@{}l@{}}TLS\\ 1.0\end{tabular} & \begin{tabular}[c]{@{}l@{}}SSL\\ 3.0\end{tabular} & \begin{tabular}[c]{@{}c@{}}Cipher \\ Suites \\ Mirroring\end{tabular} & \begin{tabular}[c]{@{}c@{}}Problematic\\ Ciphers/Hash Algorithms\end{tabular} & \begin{tabular}[c]{@{}c@{}}Key\\ Length \\ Mapping\end{tabular} & \begin{tabular}[c]{@{}c@{}}Signature\\ Algorithm \\ Mapping\end{tabular} & \begin{tabular}[c]{@{}c@{}}EV\\ Certi-\\ ficates\end{tabular} & \begin{tabular}[t]{@{}c@{}}BEAST \end{tabular} \\ \hline
\multicolumn{1}{|l|}{\textbf{Untangle}} & 1.2 & 1.2 & 1.2 & -- & \rx & 3DES & 2048 & SHA256 & DV & \rx\red{*}  \\
\multicolumn{1}{|l|}{\textbf{pfSense}} & 1.2 & -- & -- & -- & \rx &  & 2048 & SHA256 & DV &  \\
\multicolumn{1}{|l|}{\textbf{WebTitan}} & 1.2 & 1.2 & 1.2 & 1.2 & \rx & 3DES, \textbf{RC4}, IDEA  & \red{1024} & SHA256 & DV & \rx\red{*}  \\
\multicolumn{1}{|l|}{\textbf{Microsoft}} & -- & -- & 1.0 & 3.0 & \rx & 3DES, \textbf{DES}, \textbf{RC4}, \textbf{MD5} & 2048 & Mirrored & DV & \rx  \\
\multicolumn{1}{|l|}{\textbf{UserGate}} & 1.2 & 1.2 & 1.2 & -- & \rx & 3DES, \textbf{DES}  & \red{1024} & SHA256 & DV & \rx\red{*}  \\
\multicolumn{1}{|l|}{\textbf{Cisco}} & 1.2 & 1.2 & 1.2 & -- & \rx &  & \hphantom{*}2048* & SHA256 & DV & \rx   \\ 
\multicolumn{1}{|l|}{\textbf{Sophos}} & 1.2 & 1.2 & 1.2 & -- & \rx &  & 2048 & SHA256 & DV & \rx\red{*}  \\ 
\multicolumn{1}{|l|}{\textbf{TrendMicro}} & 1.2 & \textdagger & \textdagger & -- & \rx & 3DES, \textbf{RC4} & \red{1024} & SHA256 & DV & \rx   \\ 
\multicolumn{1}{|l|}{\textbf{McAfee}} & 1.2 & 1.2 & 1.2 & -- & \rx &  & 2048 & SHA256 & DV &   \\ 
\multicolumn{1}{|l|}{\textbf{Cacheguard}} & 1.2 & 1.2 & 1.2 & -- & \rx & 3DES & 2048 & SHA256 & DV & \rx\red{*} \\ 
\multicolumn{1}{|l|}{\textbf{OpenSense}} & 1.2 & 1.2 & 1.2 & -- & \gc &  & 2048 & SHA256 & DV & \rx\red{*}  \\ 
\multicolumn{1}{|l|}{\textbf{Comodo}} & 1.2 & 1.2 & 1.2 & 1.2 & \rx & 3DES, \textbf{RC4}, IDEA & 2048 & SHA256 & DV & \rx\red{*}  \\ 
\multicolumn{1}{|l|}{\textbf{Endian}} & 1.2 & 1.2 & 1.2 & 1.2 & \rx & 3DES, \textbf{RC4}, IDEA & 2048 & SHA256 & DV & \rx\red{*}   \\ \hline
\end{tabular}%
}
%\vspace{-7pt}
\end{table*}
%\vspace{-8pt}

\subsection{Private Key Protection, Self-issued, and Pre-Generated Certificates} \label{ownkey}
We attempt to locate a TLS proxy's private key (corresponding to its root certificate), and learn if it is protected adequately, e.g., inaccessible to non-root processes, encrypted under an admin password. Subsequently, we  use the located private keys to sign leaf certificates, and check if the TLS proxy accepts its own certificates as the issuing authority for externally delivered content.

To locate the private keys on the non-Windows systems, access to the network appliances' disks content and their filesystems is required. If we get access to an appliance's filesystem (following Section~\ref{catrusted}), we search for files with the following known private key file extensions: ``.pem'', ``.key'', ``.pfx'', and ``.p12'', and then compare the modulus of located RSA private keys with the proxy's public key certificate to locate the correct corresponding key. Alternatively, we locate the `squid.conf' file, the configuration file of the Squid~\cite{squid} proxy, used by most appliances as the proxy API. Squid is an open source proxy, that performs TLS interception through its `ssl\_bump' option. The configuration file points to the full path of the private key, and thus, leads us to the location of the RSA key. If the filesystem is inaccessible, we parse the raw disks for keys, using the Linux command \emph{strings} on the virtual hard disk file and search for private keys. We also use memory analysis tools, such as Volatility~\cite{foundation_2016} and Heartleech~\cite{graham_2014}, to extract the private keys in some cases; for more information, see~\cite[Appendix B]{waked}. If we acquire the private key using this methodology, we still get no information on the key's location within the appliance's file system, storage method (e.g. encrypted, obfuscated), and privileges required to access the key. For Windows-based appliances, we utilize Mimikatz~\cite{gentilkiwi_2014} to extract the private key (cf.~\cite{de2016killed}). Key storage is usually handled on Windows using two APIs: Cryptography API (CAPI), or Cryptography API: Next Generation (CNG~\cite{wienholt2007windows}). When executed with Administrator privileges, Mimikatz exports private keys that are stored using CAPI and CNG. We check the location of the private keys, the privilege required to access them and if any encryption or obfuscation is applied.
If the located private key on disk is encrypted, we rely on a python script to launch a dictionary attack. 

We also check if appliance vendors rely on pre-generated certificates for their proxies, which could be very damaging. We install two instances of the same product, and compare the certificates along with their correspondent private keys (if located). If we find the same key, we conclude that the appliance uses a pre-generated certificate, instead of per-installation keys/certificates.

\section{Results}\label{results1}
In this section, we detail the results of our analysis on TLS parameters, certificate validation, trusted certificate stores, and private key protection.

%\vspace{-8pt}
\subsection{TLS Parameters}
\label{tlspararesults}
Table~\ref{table1} shows an overview of our results.

\subhead{TLS versions and mapping} \label{tlsversionmapping}
Ten appliances support TLS versions 1.2, 1.1, and 1.0, among which three also support SSL 3.0. pfSense supports TLS 1.2 only (restricting access to some sites). Microsoft supports only TLS 1.0 and (more worryingly) SSLv3; as many web servers nowadays do not support these versions (specifically SSLv3), clients behind Microsoft will be unable to visit these websites (Over 25\% of web servers do not support TLS 1.1 \& TLS 1.2.~\cite{study}).

TrendMicro terminates the TLS connections if the highest TLS version supported by the client is not supported by the requested server, instead of using a lower TLS version that is supported by both the client and the server. For example, if the requesting client supports TLS versions 1.2, 1.1 and 1.0, and the requested server supports TLS 1.1 and 1.0 only, TrendMicro terminates the connection (with a handshake failure) instead of establishing it with the TLS version 1.1. This behavior is a more restrictive form of TLS version mirroring. Except Microsoft and TrendMicro, other appliances map all the proxy-to-server TLS versions to TLS 1.2 for the client-to-proxy connection, and thus mislead browsers/users through this artificial version upgrade.

\subhead{Certificate parameters and mapping} No appliance, except Cisco, mirrors the RSA key sizes; instead, they use a hard-coded key length for all generated certificates (i.e., artificially upgrade/downgrade the external key-length to RSA-2048, and thus may mislead clients/users). When exposed to RSA-512 and RSA-1024 server certificates, Cisco mirrors those key lengths to client-to-proxy TLS connection. However, when exposed to RSA-2048, RSA-4096 and RSA-8196, Cisco maps those key lengths to a static RSA-2048 key size for the client-to-proxy TLS connection. Three appliances use the currently non-recommended RSA-1024 certificates~\cite{barker2015nist}.
Microsoft is the only appliance which mirrors the hash algorithm; the remaining appliances use SHA256, thus making external SHA1-based certificates (considered insecure) invisible to browsers.

All appliances intercept TLS connections with EV certificates, and thus, inevitably downgrade any EV certificate to DV (as the proxies cannot generate EV certificates).

%Table 2
\begin{table*}[htb]
\centering
\caption{Results for certificate validation, part I. \rc \color{black}{} means a faulty certificate is accepted and converted to a valid certificate by the TLS proxy; \ra \color{black}{} means a faulty certificate is accepted by the TLS proxy but caught by the client browser (Firefox); and blank means certificate blocked. Endian* does not have certificate validation enabled by default.}
\label{table2}
\resizebox{\textwidth}{!}{%
\begin{tabular}{l|c|c|c|c|c|c|c|c|c|c|c|}
\cline{2-12}
 & \begin{tabular}[c]{@{}c@{}}Self\\ Signed\end{tabular} & \begin{tabular}[c]{@{}c@{}}Signature\\ Mismatch\end{tabular} & \begin{tabular}[c]{@{}c@{}}Fake\\ Geo-\\Trust\end{tabular} & \begin{tabular}[c]{@{}c@{}}Wrong\\ CN\end{tabular} & \begin{tabular}[c]{@{}c@{}}Unkn-\\own\\ Issuer\end{tabular} & \begin{tabular}[c]{@{}c@{}}Non-CA\\ Interm-\\ediate\end{tabular} & \begin{tabular}[c]{@{}c@{}}X509v1\\ Interm-\\ediate\end{tabular} & \begin{tabular}[c]{@{}c@{}}Invalid\\ pathLen-\\ Constraint\end{tabular} & \begin{tabular}[c]{@{}c@{}}Bad\\ Name\\ Constraint\\ Intermediate\end{tabular} & \begin{tabular}[c]{@{}c@{}}Unknown\\ Critical\\ X509v3\\ Extension\end{tabular} & \begin{tabular}[c]{@{}c@{}}Malformed\\ Extension\\ Values\end{tabular} \\ \hline
\multicolumn{1}{|l|}{\textbf{Untangle}} &   &   &   & \ra &   &   &   &   & \rc &   & \rc \\
\multicolumn{1}{|l|}{\textbf{pfSense}} &   &   &   &   &   &   &   &   & \rc &   & \ra  \\
\multicolumn{1}{|l|}{\textbf{WebTitan}} & \rc & \rc & \rc & \ra & \rc & \rc & \rc & \rc & \rc & \rc & \ra \\
\multicolumn{1}{|l|}{\textbf{Microsoft}} &   &   &   &   &   &   &   &   & \rc &   & \rc \\
\multicolumn{1}{|l|}{\textbf{UserGate}} & \rc & \rc & \rc & \ra & \rc & \rc & \rc & \rc & \rc & \rc & \ra \\
\multicolumn{1}{|l|}{\textbf{Cisco}} &   &   &   & \ra &   &   &   &   & \rc &   &  \\
\multicolumn{1}{|l|}{\textbf{Sophos}} &   &   &   &   &   &   &   &   & \rc &   &  \rc \\
\multicolumn{1}{|l|}{\textbf{TrendMicro}} &   &   &   &   &   &   &   &   & \rc &   &  \rc \\
\multicolumn{1}{|l|}{\textbf{McAfee}} &   &   &   &   &   &   &   &   & \rc & \rc & \rc \\
\multicolumn{1}{|l|}{\textbf{Cacheguard}} & \rc &   &   &   &   &   &   &   & \rc &   &  \ra \\
\multicolumn{1}{|l|}{\textbf{OpenSense}} &   &   &   &   &   &   &   &   & \rc &   &  \ra \\
\multicolumn{1}{|l|}{\textbf{Comodo}} & \ra & \rc & \rc & \ra & \rc & \rc & \rc & \rc & \rc & \rc & \ra \\
\multicolumn{1}{|l|}{\textbf{Endian*}} &   &    &   &   &   &   &   &   & \rc &   &  \ra \\ \hline
\end{tabular}%
}
%\vspace{-15pt}
\end{table*}

\subhead{Cipher suites} We use the Qualys Client Test~\cite{qualys2014labs} to determine the list of cipher suites used by the TLS proxies. 
Only OpenSense mirrors the client's cipher suites to the server side. Each of our test client's (Chrome/Firefox/IE) own list of cipher-suite is displayed on the Qualys test when the connection is proxied by OpenSense.

All other appliances use a hard-coded list of cipher suites instead; only four offer cipher suites that exclude any weak ciphers or hash algorithms. Eight of them offer 3DES, now considered weak due its relatively small block size~\cite{bhargavan2016practical}. Two appliances offer the insecure DES cipher~\cite{des}. Five appliances include the RC4 cipher, which has been shown to have biases~\cite{vanhoef2015all}, and is no longer supported by any modern browsers. Microsoft includes the deprecated MD5 hash algorithm~\cite{wang2005break}. Three appliances offer IDEA ciphers~\cite{biham2015new} with a 64-bit block length. When relying on the DHE ciphers, a reasonably secure modulus value should be used, e.g., 2048 or higher~\cite{adrian2015imperfect}. However, eleven appliances accept a 1024-bit modulus; UserGate and Comodo even accept a 512-bit modulus.

%Table 3

\begin{table*}[htb]
\centering
\caption{Results for certificate validation, part II. For legend, see Table~\ref{table2}; N/A means not tested as the appliance disallows adding the corresponding faulty CA certificate to its trusted store.}
\label{table3}
\resizebox{\textwidth}{!}{%
\begin{tabular}{l|c|c|c|c|c|c|c|c|c|c|c|}
\cline{2-12}
 & Revoked & \begin{tabular}[c]{@{}c@{}}Expired\\ Leaf\end{tabular} & \begin{tabular}[c]{@{}c@{}}Expired\\ Interm-\\ ediate\end{tabular} & \begin{tabular}[c]{@{}c@{}}Expired\\ Root\end{tabular} & \begin{tabular}[c]{@{}c@{}}Not Yet\\ Valid\\ Leaf\end{tabular} & \begin{tabular}[c]{@{}c@{}}Not Yet\\ Valid\\ Interm-\\ ediate\end{tabular} &\begin{tabular}[c]{@{}c@{}}Not Yet\\ Valid\\ Root\end{tabular} & \begin{tabular}[c]{@{}c@{}}Leaf \\ keyUsage\\ w/out Key\\ Enciph-\\ erment\end{tabular} & \begin{tabular}[c]{@{}c@{}}Root \\ keyUsage\\ w/out\\ KeyCert-\\ Sign\end{tabular} & \begin{tabular}[c]{@{}c@{}}Leaf\\ extKey-\\ Usage w/\\ clientAuth\end{tabular} & \begin{tabular}[c]{@{}c@{}}Root\\ extKey-\\ Usage\\ w/ Code\\ Signing\end{tabular} \\ \hline
\multicolumn{1}{|l|}{\textbf{Untangle}} & \rc &   &   & \rc &   &   & \rc &   & \rc &   & \rc \\
\multicolumn{1}{|l|}{\textbf{pfSense}} & \rc &   &   &   &   &   &   &   &   &   & \rc \\
\multicolumn{1}{|l|}{\textbf{WebTitan}} & \rc & \ra & \rc & \rc & \ra & \rc & \rc & \ra & \rc & \ra & \rc \\
\multicolumn{1}{|l|}{\textbf{Microsoft}} &   &   &   &   &   &   &   & \rc &   &   & \rc \\
\multicolumn{1}{|l|}{\textbf{UserGate}} & \rc & \rc & \rc & \rc & \rc & \rc & \rc & \rc & \rc & \rc & \rc \\
\multicolumn{1}{|l|}{\textbf{Cisco}} &   & \ra & \rc & N/A & \ra & \rc & N/A &   & N/A &   & N/A \\
\multicolumn{1}{|l|}{\textbf{Sophos}} & \rc &   &   &   &   &   &   &   &  &   &  \rc \\
\multicolumn{1}{|l|}{\textbf{TrendMicro}} & \rc &   &   &   &   &   &   &   &  &   &  \rc \\
\multicolumn{1}{|l|}{\textbf{McAfee}} &   &   &   &   &   &   &   & \rc &  & \rc & \rc \\
\multicolumn{1}{|l|}{\textbf{Cacheguard}} & \rc &   &   &   &   &   &   &   &  &   &  \rc \\
\multicolumn{1}{|l|}{\textbf{OpenSense}} & \rc &   &   &   &   &   &   &   &  &   &  \rc \\
\multicolumn{1}{|l|}{\textbf{Comodo}} & \rc & \ra & \rc & \rc & \ra & \rc & \rc & \ra & \rc & \ra & \rc \\
\multicolumn{1}{|l|}{\textbf{Endian*}} & \rc &   &   &   &   &   &   &   &  &   &  \rc \\ \hline
\end{tabular}%
}
%\vspace{-15pt}
\end{table*}

\vspace{-8pt}
\subsection{Certificate Validation Results}

In this section, we discuss the vulnerabilities found in the certificate validation mechanism of the tested TLS proxies; for summary, see Tables~\ref{table2}, \ref{table3} and \ref{table4}.

WebTitan, UserGate and Comodo do not perform \emph{any} certificate validation; their TLS proxies allowed \emph{all} our faulty TLS certificates. UserGate enables TLS inspection by default after a fresh installation. Endian does not perform certificate validation by default (a checkbox for accepting all certificates is checked by default). We uncheck the checkbox to test the certificate validation mechanism in Endian, and discuss the results based on the forced certificate validation. Comodo also includes in its configuration interface a checkbox for accepting all certificates, checked by default. Even after unchecking it, the appliance still does not perform any certificate validation. 

Both WebTitan and UserGate block access to the web servers offering RSA-512 certificates, possibly triggered by the TLS libraries utilized by the proxies, and not by the TLS interception certificate validation code (as apparent from the error messages we observed). Although Comodo accepts self-signed certificates, Firefox caught the faulty certificate. This is the result of Comodo mirroring the X.509 version 3 extension `basic constrains: CA' value of the server self-signed certificate to the client-side TLS connection. Note that Firefox blocks a TLS connection when the delivered leaf certificate has the CA flag set to true, while Chrome accepts it. We omit WebTitan, UserGate and Comodo from the remaining discussion here, as they do not perform any certificate validation.

UserGate and TrendMicro cache TLS certificates and ignore changes in the server-side certificates (as opposed to modern browsers). Therefore, we regenerate a key pair for their TLS proxies for each of our certificate validation test, to ensure accuracy (i.e., not the results of cached TLS certificates).

We mark a faulty certificate as \emph{passed} when the TLS proxy accepts the faulty certificate but leaves some chances for a diligent client to catch the anomaly. This behavior results from the way TLS proxies mirror X.509 extensions and their values to the client-to-proxy connection. The parameters mirrored are typically the common name, the keyUsage and extKeyUsage extensions, and the not before and expiry dates. In addition, Comodo mirrors the basic constraints CA flag, Cisco mirrors the RSA key size when it is 1024 bits and lower, and Microsoft mirrors the signature hashing algorithm. For simplicity, we report our results using the Firefox browser, but some results may change based on the client's validation process. For example, the Chrome browser allowed the leaf certificate with the basic constraints CA flag set to true in the Comodo self-signed test, while Firefox blocked access in this case.

Cacheguard accepts self-signed certificates (explicitly allowed in its default configuration). Untangle and Cisco forward the wrong CN certificates to our Firefox browser, which caught it and blocked access. McAfee is the sole appliance to accept a leaf certificate with an unknown x509 version 3 extension, marked as critical. Regarding malformed extension values, only Cisco blocks the anomalous certificate; pfSense, Cacheguard, OpenSense, and Endian pass it to the browser. Only Microsoft, Cisco, and McAfee check the revocation status of the offered certificates. When exposed to expired or not yet valid leaf certificates, Cisco forwards the certificates to the browser, as its default settings are configured to only monitor expired leaf certificates, and not to drop the connections. 

Untangle fails to detect expired or not yet valid root CA certificates; Cisco disallows adding them to its trusted store in the first place. Cisco fails to detect expired and not yet valid intermediate certificates. Microsoft and McAfee allow leaf certificates whose keyUsage do not include keyEnciphernment. Untangle fails to detect root CA certificates that do not have keyCertSign among the keyUsage values, and Cisco disallows adding them to its trusted store. Similarly, Cisco disallows adding root CA certificates whose extKeyUsage parameter is codeSigning and RSA-512 root CA certificates to its store. McAfee accepts leaf certificates whose extKeyUsage x509 version 3 parameter is set to clientAuth.

%Table 4

\begin{table*}[htb]
\centering
\caption{Results for certificate validation, part III. For legend, see Table~\ref{table2}.}
\label{table4}
\resizebox{.82\textwidth}{!}{%
\begin{tabular}{l|c|c|c|c|c|c|c|c|c|c|c|c|}
\cline{2-13}
 & \multicolumn{2}{c|}{\begin{tabular}[c]{@{}c@{}}Root Key \\ Length\\ (Good Leaf)\end{tabular}} & \multicolumn{4}{c|}{\begin{tabular}[c]{@{}c@{}}Leaf Key \\ Length\\ (Good Root)\end{tabular}} & \multicolumn{3}{c|}{\begin{tabular}[c]{@{}c@{}}Signature Hashing \\ Algorithm\end{tabular}} & \multicolumn{2}{c|}{\begin{tabular}[c]{@{}c@{}}DHE\\ Modulus\\ Length\end{tabular}} & \begin{tabular}[c]{@{}c@{}}Own\\ Root\end{tabular} \\ \cline{2-12} 
 & 512 & 1024 & 512 & 768 & 1016 & 1024 & MD4 & MD5 & SHA1 & 512 & 1024 &  \\ \hline
\multicolumn{1}{|l|}{\textbf{Untangle}} & \rc & \rc &   & \rc & \rc & \rc &   &  & \rc &   & \rc &  \\
\multicolumn{1}{|l|}{\textbf{pfSense}} & \rc & \rc &   & \rc & \rc & \rc &   & \rc & \rc &   & \rc &  \\
\multicolumn{1}{|l|}{\textbf{WebTitan}} & \rc & \rc &   & \rc & \rc & \rc & \rc & \rc & \rc &   & \rc & \rc \\
\multicolumn{1}{|l|}{\textbf{Microsoft}} &   & \rc &   &   &   & \rc & \ra & \ra & \ra &   & \rc & \rc \\
\multicolumn{1}{|l|}{\textbf{UserGate}} & \rc & \rc &   & \rc & \rc & \rc & \rc & \rc & \rc & \rc & \rc & \rc \\
\multicolumn{1}{|l|}{\textbf{Cisco}} & N/A & \rc & \ra & \rc & \rc & \rc & \rc & \rc & \rc &   & \rc &  \\ 
\multicolumn{1}{|l|}{\textbf{Sophos}} & \rc & \rc & \rc & \rc & \rc & \rc &   & \rc & \rc &   &  &  \\ 
\multicolumn{1}{|l|}{\textbf{TrendMicro}} & \rc & \rc &   & \rc & \rc & \rc &   & \rc & \rc &   & \rc &  \\ 
\multicolumn{1}{|l|}{\textbf{McAfee}} & \rc & \rc &   &  &  &  &   &  &  &   &  &  \\ 
\multicolumn{1}{|l|}{\textbf{Cacheguard}} & \rc & \rc & \rc & \rc & \rc & \rc &   & \rc & \rc &   & \rc &  \\ 
\multicolumn{1}{|l|}{\textbf{OpenSense}} & \rc & \rc & \rc & \rc & \rc & \rc &   & \rc & \rc &   & \rc &  \\ 
\multicolumn{1}{|l|}{\textbf{Comodo}} & \rc & \rc & \rc & \rc & \rc & \rc & \rc & \rc & \rc & \rc & \rc & \rc \\ 
\multicolumn{1}{|l|}{\textbf{Endian*}} & \rc & \rc & \rc & \rc & \rc & \rc &   & \rc & \rc &   & \rc &  \\ \hline
\end{tabular}%
}
%\vspace{-15pt}
\end{table*}

Sophos, Cacheguard, OpenSense, and Endian accept RSA-512 leaf certificates (easily factorable~\cite{valenta2016factoring}), and then issue certificates with RSA-2048, leaving no options for browsers to catch such certificates. Cisco also allows RSA-512 certificates, but Firefox detects them, as Cisco's proxy mirrors the RSA key sizes of RSA-512 and RSA-1024 server certificates to the client-to-proxy TLS connection (RSA-2048 and higher key sizes are mapped to RSA-2048).

Microsoft mirrors signature hashing algorithms, and thus passes weak and deprecated hash algorithms (if any) to the client. Cisco accepts certificates signed using the deprecated MD4 algorithm. Microsoft, WebTitan, UserGate, and Comodo fail to detect external leaf certificates signed by their own root keys.
Note that, when a TLS connection is terminated, Untangle and Microsoft use a TLS handshake failure; pfSense, Sophos, TrendMicro, McAfee, Cacheguard, OpenSense, and Endian redirect the connection to an error page; and Cisco uses an untrusted CA certificate, relying on the browser to block the connection. However, error pages as displayed by Sophos and TrendMicro, allow end-users to reestablish the connection (Sophos through `Add exception for this URL', and TrendMicro through `Continue at own your risk'). This behavior is a deviation from current practice (in browsers), as the users may be unaware of the actual risks and consequences if they bypass these warnings.

\begin{table*}[htb]
\centering
\caption{Results for trusted stores, private keys and initial setup. `N/A': not available  (failed to locate the private key on disk); `Not Applicable': the appliance does not rely on any store (no certificate validation); `World': readable by any user account on the appliance.} 
\label{table5}
\resizebox{\textwidth}{!}{%
\begin{tabular}{l|c|c|c|c|c|c|c|}
\cline{2-8}
 & \multicolumn{2}{c|}{Trusted CA Store} & \multicolumn{3}{c|}{Private Key} & \multicolumn{2}{c|}{Initial Behavior} \\ \cline{2-8} 
 & Location & Type & Location & State & Read Permission & \begin{tabular}[c]{@{}c@{}}Inspection \\ By Default\end{tabular} & \begin{tabular}[c]{@{}c@{}}Pre-Generated\\ Key Pair\end{tabular} \\ \hline
\multicolumn{1}{|l|}{\textbf{Untangle}} & \begin{tabular}[c]{@{}c@{}}/usr/share/untangle/lib/\\ssl-inspector/trusted-ca-list.jks\end{tabular} & \begin{tabular}[c]{@{}c@{}}Java \\ Key Store\end{tabular} & \begin{tabular}[c]{@{}c@{}}/usr/share/untangle\\ /settings/untangle-\\certificates/untangle.key \\ \phantom{.} \end{tabular} & Plaintext & Root & Off & No \\
\multicolumn{1}{|l|}{\textbf{pfSense}} & \begin{tabular}[c]{@{}c@{}}/usr/local/share\\ /certs/ca-root-nss.crt\end{tabular} & \begin{tabular}[c]{@{}c@{}}Mozilla\\ NSS\end{tabular} & \begin{tabular}[c]{@{}c@{}}/usr/local/etc\\ /squid/serverkey.pem \\ \phantom{.} \end{tabular} & Plaintext & World & Off & No \\
\multicolumn{1}{|l|}{\begin{tabular}[l]{@{}l@{}}\textbf{WebTitan}\end{tabular}} & Not Applicable & None & \begin{tabular}[c]{@{}c@{}}/usr/blocker/ssl\\ /ssl\_cert/webtitan.pem  \end{tabular} & Plaintext & World & Off & No \\
\multicolumn{1}{|l|}{\textbf{Microsoft}} & \begin{tabular}[c]{@{}c@{}}mmc.exe $\rightarrow$ Windows Trusted \\ Store $\rightarrow$ Local Computer\end{tabular} & \begin{tabular}[c]{@{}c@{}}Microsoft\\ Store\end{tabular} & \begin{tabular}[c] {@{}c@{}} \phantom{.} \\ CERT\_SYSTEM\_STORE\\ \_LOCAL\_MACHINE\_MY \\ \phantom{.} \end{tabular} & \begin{tabular}[c]{@{}c@{}}Exportable\\ Key\end{tabular} & Admin & Off & No \\
\multicolumn{1}{|l|}{\textbf{UserGate}} & Not Applicable & None & \begin{tabular}[c]{@{}c@{}}/opt/entensys/webfilter\\ /etc/private.pem \end{tabular} & Plaintext & World & On & No \\
\multicolumn{1}{|l|}{\begin{tabular}[l]{@{}l@{}}\textbf{Cisco}\end{tabular}} & \begin{tabular}[c]{@{}c@{}} \\ \phantom{.}Network $\rightarrow$ Certificate \\ Management $\rightarrow$ Cisco \\ Trusted Root Certificate List \\ \phantom{.} \end{tabular} & GUI & N/A & N/A & N/A & Off & No \\ 
\multicolumn{1}{|l|}{\textbf{Sophos}} & \begin{tabular}[c]{@{}c@{}}Web Protection $\rightarrow$ Filtering \\ Options $\rightarrow$ HTTPS CAs $\rightarrow$ \\ Global Verification CAs\end{tabular} & GUI & N/A & N/A & \begin{tabular}[c]{@{}c@{}}Admin (for \\ GUI download) \end{tabular} & Off & No \\ 
\multicolumn{1}{|l|}{\begin{tabular}[l]{@{}l@{}}\textbf{TrendMicro}\end{tabular}} & \begin{tabular}[c]{@{}c@{}} \\ HTTP $\rightarrow$ Configuration $\rightarrow$ \\ Digital Certificates $\rightarrow$ \\ Active Certificates \\ \phantom{.} \end{tabular} & GUI & \begin{tabular}[c]{@{}c@{}}/var/iwss/https/certstore\\ /https\_ca/default\_key.cer\end{tabular} & \begin{tabular}[c]{@{}c@{}}Passphrase \\ Encryption \end{tabular}& World & Off & \red{Yes} \\ 
\multicolumn{1}{|l|}{\textbf{McAfee}} & \begin{tabular}[c]{@{}c@{}}  Policy $\rightarrow$ Lists $\rightarrow$\\ Subscribed Lists $\rightarrow$ Certificate \\ Authorities $\rightarrow$ Known CAs \end{tabular} & GUI & N/A & N/A & \begin{tabular}[c]{@{}c@{}}Admin (for \\ GUI download) \end{tabular} & Off & \red{Yes} \\ 
\multicolumn{1}{|l|}{\begin{tabular}[l]{@{}l@{}}\textbf{Cacheguard}\end{tabular}} & \begin{tabular}[c]{@{}c@{}} \\ \phantom{.} /usr/local/proxy/var \\ /ca-ssl/ca-bundle.crt  \\ \phantom{.} \end{tabular} & \begin{tabular}[c]{@{}c@{}}Mozilla\\ NSS\end{tabular} & \begin{tabular}[c]{@{}c@{}}/usr/local/proxy/var \\ /ca-ssl/self-ca.key \end{tabular} & Plaintext & World & Off & \red{Yes} \\
\multicolumn{1}{|l|}{\textbf{OpenSense}} & /usr/local/openssl/cert.pem & \begin{tabular}[c]{@{}c@{}} Mozilla\\ NSS\end{tabular} & /var/squid/ssl/ca.pem & Plaintext & Root & Off & No \\ 
\multicolumn{1}{|l|}{\textbf{Comodo}} & Not Applicable & \begin{tabular}[c]{@{}c@{}} \\ \phantom{.} None \\ \phantom{.}\end{tabular} & /var/cni/credentials/ca.key & Plaintext & World & Off & No \\ 
\multicolumn{1}{|l|}{\begin{tabular}[l]{@{}l@{}}\textbf{Endian}\end{tabular}} & /etc/ssl/certs/ca-certificates.crt & \begin{tabular}[c]{@{}c@{}}`update-ca-\\certificates' \\Command \end{tabular} & /var/efw/proxy/https\_cert & Plaintext & World & Off & No \\ \hline
\end{tabular}%
}
%\vspace{-15pt}
\end{table*}

%\vspace{-8pt}
\subsection{Trusted CA Stores}
In this section, we analyze the results for trusted CA stores, their accessibility, source, and content; see Table~\ref{table5}.
Note that, as WebTitan, UserGate, and Comodo perform no certificate validations, their trusted stores are of no use.

\subhead{Accessing the trusted stores} Untangle's file system can be accessed through SSH. We found that Untangle relies on two CA trusted stores, saved in Java Keystore files on the filesystem. The first store, `trusted-ca-list.jks', holds the CA authorities trusted by default, while the second store, `trustStore.jks', holds the custom CA certificates, added by the machine administrator through Untangle's UI. pfSense also allows SSH, and we found that its CA trusted store on the FreeBSD filesystem under `ca-root-nss.crt'. pfSense does not offer any UI to add custom CA certificates. We append our crafted certificates to the original store, in a format that includes the public key, in addition to the text meta-data (OpenSSL's `-text' option). Microsoft relies on the Windows Server's standard trusted store. To view the content of the trusted store and to inject our crafted CA certificates, we rely on the Microsoft Management Console, in the Trusted Root Certification Authorities section of the Local Computer. 

Cisco's trusted CA store can be accessed through the appliance's web interface, under the Cisco Trusted Root Certificate List. It also includes another interface, the Cisco Blocked Certificates, for untrusted issuer certificates. To add custom CA certificates, the appliance includes a third interface, the Custom Trusted Root Certificates. However, Cisco does not allow the injection of most of our invalid root certificates, and responds with an error when tried. Sophos allows accessing the trusted CA store through its web interface, under Global Verification CAs. The interface allows adding custom root certificates, in addition to disabling CA certificates that are included by default. TrendMicro's trusted store can be accessed through the web interface's Active Certificates section. It is possible to add custom CA certificates and deactivate existing default ones. McAfee gives access to the root certificates supplied by default in the Known CAs section, and allows adding custom root certificates in the My CAs section. 

Cacheguard's web interface does not include a section for root CA certificates. In addition, Cacheguard does not give access to its filesystem through a terminal, and does not support SSH. We thus mount the appliance's virtual hard disk to a Linux machine, and locate the trusted store in a `ca-bundle.crt' file. We subsequently append our custom CA certificates to the bundle. Similarly, OpenSense and Endian do not include a section for root certificates. However, they give access to the filesystem through an OS shell terminal. We locate the trusted store of OpenSense in a `cert.pem' bundle file, and Endian's in a `ca-certificates.crt' bundle file. We include our custom CA certificates to these files.

\subhead{Source and content} As documented on Untangle's SSL Insepctor wiki page~\cite{inspector_untangle}, the list of trusted certificates is generated from the Debian/Linux ca-certificates package, in addition to Mozilla's root certificates. However, Untangle includes Microsoft's own Root Agency certificate, indicating the additional inclusion of Windows trusted certificates. The Root Agency certificate is RSA-512 that can be trivially compromised (see Section~\ref{practicalattacks}). Untangle also includes 21 RSA-1024 root certificates, 30 expired certificates, and 16 certificates from issuers that are no longer trusted by major browser/OS vendors (three from CNNIC CA, six DigiNotar, three T\"URKTRUST, and four WoSign certificates). 

pfSense's trusted CA store relies on Mozilla's NSS certificates bundle, extracted from the nss-3.30.2 version (Apr.\ 2017), with 20 untrusted certificates omitted from the bundle, as specified in the header of the trusted store. It does not include any RSA-512 or RSA-1024 certificates, and no expired certificates. However, pfSense includes two CNNIC CA certificates, and four WoSign CA certificates. 

Similar to the other Windows OSes, the Windows Server 2008 R2 also does not display the full list of trusted root certificates in its management console, and instead, only displays the root certificates of web servers already visited. We thus rely on the Microsoft Trusted Root Certificate Program~\cite{Technet} to collect the list of certificates trusted to the date of the testing. We found that the list includes two CNNIC CA certificates, two T\"URKTRUST CA certificates, two ANSSI CA certificates, and four WoSign CA certificates. Nonetheless, the acquired list does not include the RSA key sizes of the certificates, their expiry dates, or their revocation states. 

As for Cisco, we found four problematic root CA certificates from T\"URKTRUST included into the trusted store. However, the RSA key sizes are not displayed within the UI, so we could not check for RSA-512 and RSA-1024 CA certificates.

Sophos includes two CNNIC, four WoSign, and three T\"URKTRUST CA certificates; TrendMicro has a CNNIC, two T\"URKTRUST, and 30 expired certificates; and McAfee includes a CNNIC certificate. The RSA key sizes (for all three) and expiry dates (for Sophos and McAfee) of CA certificates are not displayed within their UI, and thus we could not check for these issues.

Cacheguard's trusted store is extracted from Mozilla NSS's root certificates file `certdata.txt'~\cite{mozillastore} and converted using Curl's `mk-ca-bundle.pl' version 1.27 script~\cite{mkcabundle}, as specified in the `ca-bundle.crt' trusted store file. We parse the trusted store using OpenSSL's `-text' option to extract the certificate metadata. The trusted store contains two T\"URKTRUST, four WoSign, three expired certificates; however, it is free of RSA-512 or RSA-1024 certificates.
OpenSense's store also relies on Mozilla, extracted from the nss-3.35 (Jan.\ 2018) version, with two untrusted certificates omitted from the bundle, as specified in the header of the NSS trusted store. It does not include any RSA-512 or RSA-1024 certificates, and no expired certificates. However, it includes a T\"URKTRUST CA certificate.

Endian's trusted CA store bundle is the output of the `update-ca-certificates' Debian Linux command~\cite{cacert}. The trusted store contains two CNNIC, three T\"URKTRUST, four WoSign, 10 expired, and 11 RSA-1024 CA certificates.

%\vspace{-8pt}
\subsection{Private Key Protection}

In this section, we discuss the results regarding the TLS proxies' private keys, in terms of storage location, state, and the privilege required to access them; see Table~\ref{table5}.

We could not access the filesystem of Cisco's AsyncOS to locate its private key on disk. Instead, we extract the key from memory using Heartleech~\cite{graham_2014} (see~\cite[Appendix B]{waked}). Sophos and McAfee provide access to their filesystems through a bash terminal. However, we could not locate their private keys on disk. Sophos stores the key in a database, as it can be recovered by invoking the following command `cc get\_object REF\_CaSigProxyCa' via Sophos' terminal. McAfee possibly has its private key hard-coded, as its key pair is pre-generated, as discussed later in this section. Thus, we could not locate the private key on disk. We get a copy of their respective private keys by downloading them from the appliances' web interfaces. As there is no information on the private key on disk, and the located key was used only for testing external content signed by own key, we do not discuss these appliances in the rest of the section.

We rely on the methodologies from Section~\ref{catrusted} to access the filesystems on non-Windows appliances. pfSense and Untangle provide SSH access. For WebTitan and Cacheguard, we mount their respective virtual disk disks on a separate machine. UserGate, TrendMicro, OpenSense, Comodo, and Endian provide access to their OS shell terminal by default. Untangle, pfSense, WebTitan, UserGate, Cacheguard, OpenSense, Comodo and Endian store their plaintext private keys within their filesystems (as `untangle.key', `serverkey.pem', `webtitan.pem', `private.pem', `self-ca.key', `ca.pem', `ca.key', and `https\_cert' files, respectively). pfSense, WebTitan, UserGate, Cacheguard, Comodo, and Endian allow read access to all users accounts (write is restricted to root), while Untangle and OpenSense allow read/write only to root accounts.

Regarding TrendMicro, we get access to the filesystem using its OS terminal, and locate the root private key in a file named `default\_key', with read permission to all user accounts (write is restricted to root). However, the located key is encrypted using a passphrase. We brute-force the encrypted key using a python script and a dictionary of common English words, and successfully decrypt the key, with the passphrase `trend'.

Microsoft's private key is stored using the Windows Software Key Storage Provider, utilizing Cryptography API: Next Generation (CNG). The key is exportable through the Microsoft Management Console, if opened with SYSTEM privileges. We rely on the Mimikatz tool to export the key, which requires a less privileged Administrator account. 

We install multiple instances of each appliance to check if the root certificates are pre-generated. To our surprise, we found that TrendMicro, McAfee and Cacheguard use such certificates to intercept the TLS traffic. McAfee includes an X509v3 `Netscape Certificate Comment' extension, with the following warning: ``This is the default McAfee root CA. It will be delivered with each web gateway installation. We recommend to generate and use your own CA.''. However, it does not provide any warning during installation/configuration. Although Cacheguard's documentation explicitly state: ``the default system CA certificate is generated during the installation''~\cite{cacheguardCA}, in reality, it uses a pre-generated certificate.

%\vspace{-8pt}
\section{Evolution of Products Between 2016--2018}
\label{differences}

In this section, we highlight the evolution of the overlapping appliances that were tested in three separate instances between 2016 to 2018: by Durumeric et al.~\cite{durumeric2017security} in 2016 (disclosed to vendors in Sept.\ 2016), our own tests in 2017~\cite{waked} (disclosed in Dec.\ 2017), and the latest product releases tested in 2018 as part of this paper (disclosed in May 2018).

In 2016, Untangle included RC4 and weak ciphers in its cipher-suite; we found that version 13.0 (2017) still included weak ciphers, but no RC4. The Untangle 13.2 release, tested in 2018, has no differences in its TLS interception processes compared to release 13.0, and thus, shows the exact same results. pfSense, which was not tested in 2016 by Durumeric et al., accepts the TLS version 1.1 in its 2.3.4 release (2017), while pfSense 2.4.2-P1 (2018) no longer does. Moreover, pfSense 2.3.4 maps the certificate keys to RSA-4096, while the latest version maps them to RSA-2048. In 2016, WebTitan had a broken certificate validation process and offered RC4 and modern ciphers; we found that WebTitan version 5.15 (2017) did not perform \emph{any} certificate validation, was vulnerable to the CRIME attack, and still offered RC4, in addition to weak ciphers. Moreover, the latest version of WebTitan (5.16) in 2018 accepts SSLv3 (did not in 2017), but is now patched against CRIME. Microsoft performed no certificate validation in 2016 and the highest supported SSL/TLS version was SSLv2.0; it now (2018) performs certificate validation, and supports SSL versions 2.0, 3.0 and TLS 1.0. The Microsoft and UserGate product releases are the same in 2018 compared to 2017. Cisco no longer offers RC4 and export-grade ciphers, which was reported in 2016. Furthermore, Cisco build 270's CBC ciphers (2017) are not recognized by the Qualys client test, while the latest build's CBC ciphers (2018) are, indicating that the appliance is vulnerable to the BEAST attack. The older build fails to block RSA certificates with malformed extension values, while the latest build does. The latest build fails to block expired and not yet valid intermediate root certificates, in addition to RSA-512 leaf certificates, while the older build (270) blocks them successfully. In 2016, Sophos offered RC4, but not in the 2018 release.

We contacted the six affected companies after our 2017 tests, and received replies from three companies; Untangle replied with just an automatic reply, Entensys confirmed that they have passed the matter to its research team. Netgate (pfSense), stated that they philosophically oppose TLS interception, but include it as it is a commonly requested feature. Netgate also states that the TLS interception is done using the external package `Squid', which it does not control completely. They claimed that our tested version was five releases old at that time. We found the latest version to have the exact same results, with two minor exceptions. We are also contacting all vendors from our latest 2018 tests.

Overall, the disclosures appear to have limited impact on vendors. Many vendors completely ignored the security issues (Untangle, Microsoft, UserGate, and pfSense). More worryingly, some products even became worse over time (Cisco), and some patched product releases introduced new vulnerabilities compared to their older versions (WebTitan).

%\vspace{-8pt}
\section{Practical Attacks}
\label{practicalattacks}
In this section, we summarize how the vulnerabilities reported could be exploited by an attacker. 

MITM attacks can be trivially launched to impersonate any web server against clients behind UserGate, WebTitan, Comodo and Endian, due to their lack of certificate validation (using default configuration). Attackers can simply use a self-signed certificate for any desired domain, fooling even the most secure and up-to-date browsers behind these appliances. Since Usergate enables TLS interception by default, users located behind a freshly installed UserGate appliance are automatically vulnerable. Likewise, clients behind Cacheguard are vulnerable, as the appliance's TLS proxy accepts self-signed certificates. Clients behind Untangle are also similarly vulnerable, due to the RSA-512 `Root Agency' certificate in its trusted store. This Root Agency CA certificate, which is valid until 2039, has been used since the 1990s as the default test certificate for code signing and development; Windows systems still include this certificate, but mark it as untrusted. The RSA-512 private key corresponding to this certificate can be easily factored under four hours~\cite{valenta2016factoring} as a one-time effort, and the factored key could be used attack all instances of Untangle.

An attacker can also launch MITM attacks to decrypt traffic or impersonate any web server against clients behind TrendMicro, McAfee and Cacheguard, as they rely on pre-generated root keys (identical on all installations). The attacker can retrieve private keys for these appliances from her own installations irrespective of \mbox{privileges required to access the keys.} 

UserGate, WebTitan, Microsoft, and Comodo accept external certificates signed by their own root keys. If an attacker can gain access to the private keys of these appliances, she can launch MITM attacks to impersonate any web server. UserGate, WebTitan and Comodo provide `read' privileges to non-root users for the private key, while Microsoft mandates admin privileges.

When combined with a Java applet to bypass the same origin policy, the BEAST vulnerability~\cite{duong2011here} may allow an attacker to recover authentication cookies from the clients behind Microsoft, Cisco and TrendMicro. Attackers could also recover cookies from clients behind WebTitan, Microsoft, TrendMicro, Comodo, and Endian due to their use of RC4~\cite{vanhoef2015all}.

Attackers could break session confidentiality for clients behind  Sophos, Cacheguard, OpenSense, Comodo and Endian, as they accept RSA-512 external leaf certificates (RSA-512 is easily factored). Note that, in 2016, 1\% of TLS web servers were found to host an RSA-512 certificate~\cite{valenta2016factoring}. In contrast, modern browsers will refuse to establish such connections.

All appliances except Untangle and McAfee accept certificates signed using MD5, with WebTitan, Microsoft, UserGate, Cisco and Comodo also accept MD4. Weak collision resistance of MD5/MD4~\cite{wang2005break} can be exploited in a practical attack scenario, where the attacker forges a rogue intermediate CA certificate that appears to be signed by a valid trusted root CA; all leaf certificates signed by this rogue CA will similarly be trusted by the appliances. As a result of this one-time effort, the holder of this rogue intermediate CA can launch MITM attacks and impersonate web servers, targeting the users behind all the appliances that accept certificates signed using MD5~\cite{sotirov2008md5}.

%\vspace{-8pt}
\section{Conclusion}

We present a framework for analyzing TLS interception behaviors of network appliances to uncover potential vulnerabilities introduced by them. We tested thirteen network appliances, and found that all their TLS proxies are vulnerable against the tests under our framework---at varying levels. Each proxy lacks at least one of the best practices in terms of protocol and parameters mapping, patching against known attacks, certificate validation, CA trusted store maintenance, and private key protection.
We found that the clients behind the thirteen appliances are vulnerable to full server impersonation under an active MITM attack, of which one enables TLS interception by default. We also found that three TLS proxies rely on pre-generated root keys, allowing trivial MITM attacks. 

While TLS proxies are mainly deployed in enterprise environments to decrypt the traffic in order to scan for malware and network attacks, they introduce new intrusion opportunities and vulnerabilities for attackers. As TLS proxies act as the client for the proxy-to-web server connections, they should maintain (at least) the same level of security as modern browsers; similarly, as they act as a TLS server for the client-to-proxy connections, they should be securely configured like any up-to-date HTTPS server, by default. Before enabling TLS interception, concerned administrators may use our framework to evaluate their network appliances, and weigh the potential vulnerabilities that may be introduced by a TLS proxy against its perceived benefits.

\section*{Acknowledgements}
We thank the ASIA CCS 2018 anonymous reviewers for their helpful comments and suggestions to improve the paper's presentation, and Xavier de Carn\'{e} de Carnavalet for his help during the development of our framework. This research is supported by NSERC. 

\bibliographystyle{abbrv}

\begin{thebibliography}{10}

\bibitem{su_2011}
{BEAST} attack 1/n-1 split patch.
\newblock \url{https://bugzilla.mozilla.org/show_bug.cgi?id=665814#c59}, Jul
  2017.

\bibitem{cacheguardCA}
{CacheGuard OS user's guide - SSL mediation}.
\newblock \url{https://www.cacheguard.net/doc/guide/ssl_mediation.html}, Jan
  2018.

\bibitem{mkcabundle}
Curl's mk-ca-bundle.pl - {GitHub}.
\newblock \url{https://github.com/curl/curl/blob/master/lib/mk-ca-bundle.pl},
  Jan 2018.

\bibitem{prins2011diginotar}
{DigiNotar CA breach}.
\newblock
  \url{https://nakedsecurity.sophos.com/2011/09/05/operation-black-tulip-fox-its-report-on-the-diginotar-breach/}.

\bibitem{CNNIC}
Distrusting new {CNNIC} certificates.
\newblock
  \url{https://blog.mozilla.org/security/2015/04/02/distrusting-new-cnnic-certificates/},
  Apr 2015.

\bibitem{wosign_smartcom}
Distrusting new {WoSign} and {StartCom} certificates.
\newblock
  \url{https://blog.mozilla.org/security/2016/10/24/distrusting-new-wosign-and-startcom-certificates/},
  Oct 2016.

\bibitem{WDormann2}
Effects of {HTTPS} and {SSL} inspection on the client.
\newblock
  \url{https://vuls.cert.org/confluence/display/Wiki/Effects+of+HTTPS+and+SSL+inspection+on+the+client},
  Aug 2017.

\bibitem{wilson}
Extended validation {OID}.
\newblock \url{https://cabforum.org/object-registry/}.

\bibitem{graham_2014}
Heartleech - {GitHub}.
\newblock \url{https://github.com/robertdavidgraham/heartleech}.

\bibitem{jmhodges_2013}
{Howsmyssl - GitHub}.
\newblock \url{https://github.com/jmhodges/howsmyssl}.

\bibitem{rosenblatt2015lenovo}
Lenovo's superfish security.
\newblock
  \url{https://www.cnet.com/news/superfish-torments-lenovo-owners-with-more-than-adware/},
  Feb 2015.

\bibitem{study}
Mapping the current state of {SSL/TLS} - thesis.
\newblock
  \url{http://www.diva-portal.org/smash/get/diva2:1109739/FULLTEXT01.pdf},
  2017.

\bibitem{TMG}
Microsoft {TMG} 2010 updates.
\newblock
  \url{https://blogs.technet.microsoft.com/keithab/2011/09/27/forefront-tmg-2010-service-pack-rollup-and-version-number-reference/}.

\bibitem{TMGWind2008}
Microsoft {TMG} supported {OS} version.
\newblock \url{https://www.microsoft.com/en-ca/download/details.aspx?id=14238}.

\bibitem{Technet}
Microsoft trusted root certificate program.
\newblock
  \url{https://gallery.technet.microsoft.com/Trusted-Root-Certificate-123665ca}.

\bibitem{gentilkiwi_2014}
Mimikatz - {GitHub}.
\newblock \url{https://github.com/gentilkiwi/mimikatz}.

\bibitem{mozillastore}
Mozilla's `certdata.txt' file.
\newblock
  \url{https://hg.mozilla.org/mozilla-central/raw-file/tip/security/nss/lib/ckfw/builtins/certdata.txt},
  Sep 2017.

\bibitem{ANSSI}
Revoking {ANSSI CA}.
\newblock
  \url{https://security.googleblog.com/2013/12/further-improving-digital-certificate.html},
  Dec 2013.

\bibitem{WDormann}
The risks of {SSL} inspection.
\newblock
  \url{https://insights.sei.cmu.edu/cert/2015/03/the-risks-of-ssl-inspection.html},
  Mar 2015.

\bibitem{squid}
{SSL Bump configuration - Squid}.
\newblock \url{http://www.squid-cache.org/Doc/config/ssl_bump/}.

\bibitem{qualys2014labs}
{SSL} client test.
\newblock \url{https://www.ssllabs.com/ssltest/viewMyClient.html}.

\bibitem{ducklin2013turktrust}
The {T\"URKTRUST SSL} certificate fiasco.
\newblock
  \url{https://nakedsecurity.sophos.com/2013/01/08/the-turktrust-ssl-certificate-fiasco-what-happened-and-what-happens-next/},
  Jan 2013.

\bibitem{inspector_untangle}
Untangle {SSL} inspector documentation.
\newblock
  \url{https://wiki.untangle.com/index.php/SSL_Inspector#Trust_All_Server_Certificates}.

\bibitem{cacert}
{update-ca-certificates - Debian System Manager's Manual}.
\newblock
  \url{https://manpages.debian.org/jessie/ca-certificates/update-ca-certificates.8.en.html},
  Apr 2017.

\bibitem{foundation_2016}
Volatility.
\newblock \url{http://www.volatilityfoundation.org/26}.

\bibitem{wienholt2007windows}
Windows cryptography {API} {(CNG)}.
\newblock
  \url{https://www.codeguru.com/cpp/w-p/vista/article.php/c13813/Windows-Cryptography-API-Next-Generation-CNG.htm}.

\bibitem{adrian2015imperfect}
D.~Adrian, K.~Bhargavan, Z.~Durumeric, P.~Gaudry, M.~Green, J.~A. Halderman,
  N.~Heninger, D.~Springall, E.~Thom{\'e}, L.~Valenta, et~al.
\newblock Imperfect forward secrecy: How diffie-hellman fails in practice.
\newblock In {\em ACM CCS}, Denver, CO, USA, 2015.

\bibitem{barker2015nist}
E.~Barker and A.~Roginsky.
\newblock {NIST} recommendations.
\newblock {\em NIST Special Publication}, 800(131A):1--29, 2015.

\bibitem{durumeric2015tracking}
B.~Beurdouche, K.~Bhargavan, A.~Delignat-Lavaud, C.~Fournet, M.~Kohlweiss,
  A.~Pironti, P.-Y. Strub, and J.~K. Zinzindohoue.
\newblock A messy state of the union: Taming the composite state machines of
  tls.
\newblock In {\em IEEE Symposium on Security and Privacy}, Fairmont, CA, USA,
  2015.

\bibitem{bhargavan2016practical}
K.~Bhargavan and G.~Leurent.
\newblock On the practical (in-) security of 64-bit block ciphers: Collision
  attacks on {HTTP over TLS and OpenVPN}.
\newblock In {\em ACM CCS}, Vienna, Austria, 2016.

\bibitem{biham2015new}
E.~Biham, O.~Dunkelman, N.~Keller, and A.~Shamir.
\newblock New attacks on {IDEA} with at least 6 rounds.
\newblock {\em Journal of Cryptology}, 28(2):209--239, 2015.

\bibitem{chausymcerts}
S.~Y. Chau, O.~Chowdhury, E.~Hoque, H.~Ge, A.~Kate, C.~Nita-Rotaru, and N.~Li.
\newblock {SymCerts}: Practical symbolic execution for exposing noncompliance
  in {X.509} certificate validation implementations.
\newblock In {\em IEEE Symposium on Security and Privacy}, Fairmont, CA, USA,
  2017.

\bibitem{de2016killed}
X.~de~Carn\'{e}~de Carnavalet and M.~Mannan.
\newblock Killed by proxy: Analyzing client-end tls interception software.
\newblock In {\em NDSS'16}, San Diego, CA, USA.

\bibitem{duong2011here}
T.~Duong and J.~Rizzo.
\newblock Here come the $\oplus$ ninjas.
\newblock {\em Technical Report}.
\newblock \url{http://www.hpcc.ecs.soton.ac.uk/~dan/talks/bullrun/Beast.pdf},
  May 2011.

\bibitem{duong2012crime}
T.~Duong and J.~Rizzo.
\newblock The {CRIME} attack.
\newblock {\em Presentation at Ekoparty Security Conference}, 2012.

\bibitem{durumeric2017security}
Z.~Durumeric, Z.~Ma, D.~Springall, R.~Barnes, N.~Sullivan, E.~Bursztein,
  M.~Bailey, J.~A. Halderman, and V.~Paxson.
\newblock The security impact of {HTTPS} interception.
\newblock In {\em NDSS'17}, San Diego, CA, USA.

\bibitem{housley2008rfc}
R.~Housley, W.~Ford, W.~Polk, and D.~Solo.
\newblock {RFC} 5280: Internet x.509 public key infrastructure certificate and
  crl profile, May 2008.

\bibitem{jarmoc2012ssl}
J.~Jarmoc.
\newblock {SSL/TLS} interception proxies and transitive trust.
\newblock {\em Black Hat Europe}, Mar 2012.

\bibitem{o2016tls}
M.~O'Neill, S.~Ruoti, K.~Seamons, and D.~Zappala.
\newblock {TLS} proxies: Friend or foe?
\newblock In {\em ACM IMC'16}, Santa Monica, CA, USA.

\bibitem{rescorla2010rfc}
E.~Rescorla, M.~Ray, S.~Dispensa, and N.~Oskov.
\newblock {RFC} 5746: Transport layer security ({TLS}) renegotiation indication
  extension, Feb 2010.

\bibitem{ruoti2016user}
S.~Ruoti, M.~O'Neill, D.~Zappala, and K.~E. Seamons.
\newblock User attitudes toward the inspection of encrypted traffic.
\newblock In {\em USENIX SOUPS'16}.

\bibitem{sotirov2008md5}
A.~Sotirov, M.~Stevens, J.~Appelbaum, A.~K. Lenstra, D.~Molnar, D.~A. Osvik,
  and B.~de~Weger.
\newblock {MD5} considered harmful today, creating a rogue {CA} certificate.
\newblock In {\em Chaos Communication Congress}, 2008.

\bibitem{valenta2016factoring}
L.~Valenta, S.~Cohney, A.~Liao, J.~Fried, S.~Bodduluri, and N.~Heninger.
\newblock Factoring as a service.
\newblock In {\em FC'16}, Barbados.

\bibitem{des}
P.~Van De~Zande.
\newblock The day {DES} died.
\newblock {\em SANS Institute}, Jul 2001.

\bibitem{vanhoef2015all}
M.~Vanhoef and F.~Piessens.
\newblock All your biases belong to us: Breaking {RC4} in {WPA-TKIP} and {TLS}.
\newblock In {\em USENIX Security Symposium}, pages 97--112, Washington D.C.,
  USA, 2015.

\bibitem{waked}
L.~Waked, M.~Mannan, and A.~Youssef.
\newblock To intercept or not to intercept: Analyzing {TLS} interception in
  network appliances.
\newblock In {\em ACM ASIACCS'18}, Incheon, Korea.

\bibitem{wang2005break}
X.~Wang and H.~Yu.
\newblock How to break {MD5} and other hash functions.
\newblock In {\em Eurocrypt'05}, Aarhus, Denmark.

\end{thebibliography}

\balance
\end{document}